\documentclass[reqno,12pt]{amsart}
\usepackage{fullpage}
\usepackage{amsfonts}
\usepackage{amssymb}
\usepackage{latexsym}
\usepackage{mathrsfs}
\numberwithin{equation}{section}

\newtheorem{theorem}{Theorem}{\bf}{\it}

\newtheorem{lemma}[theorem]{Lemma}{\bf}{\it}

{\bf}{\it}

\newtheorem{proposition}[theorem]{Proposition}{\bf}{\it}

\def\p{\partial}

\def\mbf{\boldsymbol}
\def\fr{\frac}
\def\tfr{\tfrac}

\def\eos/{equation of state}
\def\esos/{equations of state}
\def\com/{constant of motion}
\def\csom/{constants of motion}
\def\ie/{i.e.}
\def\eg/{e.g.}

\def\id{{\rm id}}

\def\V{\mathcal{V}}
\def\S{\mathcal{S}}
\def\C{\mathcal{C}}

\def\grad{\vec{\nabla}}
\def\div{{\it div}}
\def\Div{{\it Div}}
\def\gnabla{\,{}^g\nabla}
\def\Dt{\mathfrak{D}_t}
\def\Lu{\mathcal{L}_u}
\def\hook{\rfloor}
\def\nor{\hat\nu}
\def\d{\mathbf{d}}

\def\u{\mathbf{u}}
\def\w{\boldsymbol{\omega}}
\def\vortvec{\vartheta}
\def\vortscal{\varpi}
\def\circ{\varGamma}

\def\e#1{{\hat e}_{(#1)}}

\def\Rnum{\mathbb{R}}
\def\const{\text{const.}}

\begin{document}

\title{New conserved vorticity integrals for moving surfaces in multi-dimensional fluid flow}

\author{
Stephen C. Anco \\
Department of Mathematics\\
Brock University, 
St. Catharines, ON Canada
}

\email{sanco@brocku.ca}

\thanks{S.C.A. is supported by an NSERC research grant.}

\begin{abstract}
For inviscid fluid flow in any $n$-dimensional Riemannian manifold, 
new conserved vorticity integrals generalizing 
helicity, enstrophy, and entropy circulation 
are derived for lower-dimensional surfaces that move along fluid streamlines. 
Conditions are determined for which the integrals yield constants of motion 
for the fluid. 
In the case when an inviscid fluid is isentropic, 
these new constants of motion generalize Kelvin's circulation theorem
from closed loops to closed surfaces of any dimension. 
\end{abstract}
\maketitle

\section{Introduction}

Vorticity conservation laws have long been of interest in the study of 
inviscid fluid flow 
(\eg/ \cite{vorticity1,vorticity2,ArnKhe}). 
Their mathematical formulation in multi-dimensions is given by 
an integral continuity equation 
\begin{equation}
\fr{d}{dt}\int_{\V(t)} T dV
= -\int_{\p\V(t)}\vec{X} \cdot d\vec{A}
\label{continuity}
\end{equation}
for a conserved density $T$ and a spatial flux $\vec{X}$ that 
have an essential dependence on the curl of the fluid velocity, 
as defined on domains $\V(t)$ that are transported 
along the streamlines of the fluid governed by Euler's equations. 
Physically speaking, such conservation laws express 
basic rotational properties of the fluid in a reference frame that moves 
with the fluid flow \cite{Ibr}.
To-date the only known vorticity conservation laws consist of 
the general enstrophy integral \cite{MajBer,KheChe,AncDar1}
which exists in two (and higher-even) spatial dimensions,
the helicity integral \cite{Mof,KheChe,AncDar1}
which exists in three (and higher-odd) spatial dimensions,
and the entropy circulation integrals \cite{Kur,AncDar2}
which exist in two as well as three (and higher) spatial dimensions. 
A general formulation of these conservation laws, 
applicable to incompressible as well as compressible inviscid fluids, 
is hard to find in the literature. 

The Eulerian fluid equations in $\Rnum^n$ are given in terms of 
the velocity $\vec{u}$, the mass density $\rho$, the entropy density $S$,
and the pressure $p$ by 
\begin{align}
& \vec{u}_t + \vec{u}\cdot\grad\vec{u} = -\rho^{-1}\grad p , 
\label{Eulervel}\\\
& \rho_t + \grad\cdot(\rho \vec{u}) = 0 , 
\label{Eulerdens}\\
& S_t + \vec{u}\cdot\grad S =0 .
\label{Eulerentr}
\end{align}
In the case of compressible fluid flow, 
$p$ is specified by an \eos/ in terms of a function $p=P(\rho,S)$,
whereas in the case of incompressible fluid flow,
$\rho$ and $S$ are constant while $p$ satisfies the Laplacian equation 
$-\rho^{-1} \Delta p = (\grad\vec{u})\cdot(\grad\vec{u})$. 

For inviscid fluid flow in $\Rnum^2$, 
enstrophy and entropy circulation are defined by 
the respective two-dimensional integrals 
\begin{equation} \int_{\V(t)} \rho^{-1}\vortscal^2\; d^2x
\end{equation}
and 
\begin{equation}
\int_{\V(t)} S\vortscal\; d^2x
= \oint_{\p\V(t)} S\vec{u}\cdot d\vec{s}+ \int_{\V(t)} \vec{u}\cdot *\grad S\; d^2x
\label{2dentropycirc}
\end{equation}
where $\vortscal=\grad\cdot*\vec{u}$ is the vorticity scalar,
given in terms of the infinitesimal rotation operator $*$. 
Enstrophy physically measures the bulk rotation of the fluid,
while entropy circulation measures the flux of entropy around closed loops 
when the streamlines are aligned with the entropy gradient in the fluid.
The more general enstrophy integral 
\begin{equation}
\int_{\V(t)} f(\vortscal/\rho) \rho\; d^2x
= \int_{\V(t)} \tilde f(\vortscal/\rho) \vortscal\; d^2x
\label{2denstrophy}
\end{equation}
is connected with circulation properties of the fluid,
when the function $f$ is odd ($\tilde f$ is even), 
and with bulk rotation moments of the fluid, 
when the function $f$ is even ($\tilde f$ is odd). 

In contrast, 
for inviscid fluid flow in $\Rnum^3$, 
helicity and entropy circulation are respectively defined by 
the three-dimensional integrals 
\begin{equation}
\int_{\V(t)} \vec{u}\cdot\vec{\vortvec}\; d^3x
\label{3dhelicity}
\end{equation}
and
\begin{equation}
\int_{\V(t)} \grad S\cdot\vec{\vortvec}\; d^3x
= \oint_{\p\V(t)} S\vec{\vortvec}\cdot d\vec{A}
= \oint_{\p\V(t)} (\vec{u}\times \grad S) \cdot d\vec{A}
\end{equation}
involving the vorticity vector $\vec{\vortvec}=\grad\times\vec{u}$, 
given in terms of the cross-product operator. 
Physically, helicity measures the net rotation of the fluid 
around the directions of streamlines,
while entropy circulation detects the net alignment between 
the streamlines and the entropy gradient. 
A more general integral 
\begin{equation}
\int_{\V(t)} f(\grad S\cdot\vec{\vortvec}/\rho)\rho\; d^3x
= \int_{\V(t)} \tilde f(\grad S\cdot\vec{\vortvec}/\rho)\grad S\cdot\vec{\vortvec}\; d^3x
\label{3dentropycirc}
\end{equation}
measures entropy circulation properties of the fluid, 
when the function $f$ is odd ($\tilde f$ is even), 
and bulk alignment between the streamlines and the entropy gradient 
in the fluid, when the function $f$ is even ($\tilde f$ is odd).

The helicity integral \eqref{3dhelicity} 
and the general enstrophy integral \eqref{2denstrophy}
are conserved whenever the fluid flow is isentropic 
(\ie/ $S$ is constant throughout the fluid domain)
and either incompressible 
(\ie/ $\rho$ is constant throughout the fluid domain)
or compressible with a barotropic \eos/ 
in which the pressure $p$ is a function only of the density $\rho$,
whereas the entropy circulation integrals \eqref{2dentropycirc} and \eqref{3dentropycirc}
are non-trivially conserved when the fluid flow is 
non-isentropic (\ie/ $S$ is constant only along streamlines)
as well as compressible (so that $\rho$ is non-constant in the fluid domain). 
All of these conserved vorticity integrals have a mathematical origin 
as Casimir invariants of the Hamiltonian formulation 
\cite{Dez,Ser,Kur,Ver,Kup}
for the underlying fluid equations \eqref{Eulervel}--\eqref{Eulerentr}.

In the present paper, 
new conserved vorticity integrals generalizing 
helicity, enstrophy, and entropy circulation 
are derived for $p$-dimensional surfaces 
that move along streamlines of inviscid fluid flow 
in any $n$-dimensional Riemannian manifold with $n>p>1$. 
In the case of odd-dimensional surfaces, 
the helicity integral yields a \com/ if the surface is boundaryless 
or satisfies a zero-vorticity boundary condition, 
while the entropy circulation integral yields a \com/ 
whether the surface is boundaryless or has a boundary. 
Similarly, in the case of even-dimensional surfaces, 
the entropy circulation integral yields a \com/ 
if the surface is boundaryless 
or satisfies a vorticity boundary condition 
involving the entropy gradient vector, 
while the enstrophy integral yields a \com/ 
whether the surface is boundaryless or has a boundary. 
As a by-product of these results, 
a simple proof of Kelvin's circulation theorem 
is obtained for isentropic fluid flow in all dimensions $n>1$,
and its relationship to the new helicity and enstrophy integrals for
two-dimensional moving surfaces is explained.

\section{Inviscid fluid equations in Riemannian manifolds}

Consider an $n$-dimensional manifold $M$ with a Riemannian metric $g$. 
Let $\nabla$ be the metric-compatible covariant derivative
determined by $\nabla g=0$, 
and let $\epsilon_g$ be the metric-normalized volume tensor 
determined by $\nabla\epsilon_g=0$ and $g(\epsilon_g,\epsilon_g)=n!$. 
Write $\gnabla$ and $\div$ 
for the vector derivative operator and the covariant divergence operator 
defined by $\xi\hook\nabla = g(\xi,\gnabla)$ 
and $g(\gnabla,\xi)=\div\xi$
holding for an arbitrary vector field $\xi$ on $M$. 
These operators are the natural covariant counterparts of 
the gradient $\grad$ and divergence $\grad\cdot$ operators in $\Rnum^n$. 

In this tensorial notation, 
the covariant generalization of Euler's equation \eqref{Eulervel}
from $\Rnum^n$ to $M$ is given by 
\begin{equation}
u_t+(u\hook\nabla) u = -\rho^{-1}\gnabla p  
\label{veleqn}
\end{equation}
where $u$ is the fluid velocity vector on $M$. 
Similarly the covariant equations for 
the fluid pressure $p$, mass density $\rho$, and entropy density $S$ on $M$
are given by 
\begin{align}
& p=P(\rho,S) , 
\label{eoseqn}\\
& \rho_t + \div(\rho u) = 0 , 
\label{denseqn}\\
& S_t + u\hook\nabla S =0 ,
\label{entreqn}
\end{align}
in the case of compressible fluid flow, 
or by 
\begin{align}
& \rho =\const, \quad S=\const, 
\label{incompreqn}\\
& -\rho^{-1} \Delta_g p = g(\gnabla u,\gnabla u) + R(u,u)
\label{presseqn}\end{align}
in the case of incompressible fluid flow,
where 
$\Delta_g=g(\gnabla,\gnabla)=\gnabla\hook\nabla$ denotes the scalar Laplacian, 
and $R$ denotes the Ricci tensor of $g$. 
In both cases, the curl of the velocity is an antisymmetric tensor on $M$
\begin{equation}
\omega=\gnabla\wedge u
\end{equation}
obeying the transport equation
\begin{equation}
\omega_t +\gnabla\wedge g(u,\omega) = \rho^{-2}\gnabla\rho\wedge\gnabla p, 
\quad
\gnabla\wedge\omega=0 .
\label{curleqn}
\end{equation}

It will be very advantageous to rewrite the fluid equations \eqref{veleqn}--\eqref{curleqn} 
by means of differential forms 
combined with the material (advective) derivative 
\begin{equation}
\Dt=D_t +\Lu . 
\end{equation}
Here $D_t$ denotes $\p_t$ acting as a total time derivative,
and $\Lu$ denotes the Lie derivative acting on $p$-forms $\mbf\alpha$ on $M$
by $\Lu \mbf\alpha= u\hook \d\mbf\alpha + \d(u\hook \mbf\alpha)$. 
Note that the exterior derivative $\d$ commutes with the Lie derivative $\Lu$
and acts as a total spatial derivative. 
Now let $\u$ be the velocity $1$-form and let $\w$ be the curl $2$-form 
defined by the respective duals of $u$ and $\omega$ 
with respect to the metric $g$ on $M$,
namely $\xi\hook\u=g(\xi,u)$ and $\zeta\hook\w=g(\zeta,\omega)$
for an arbitrary vector field $\xi$ 
and an arbitrary antisymmetric tensor field $\zeta$ on $M$. 
The Lie derivatives of $\u$ and $\w$ along streamlines are, respectively, 
\begin{equation}
\Lu\u =u\hook\w + \d(u\hook\u), 
\quad
\Lu\w =\d(u\hook\w)
\end{equation}
where
\begin{equation}
u\hook\w = (u\hook\nabla)\u -\d(\tfr{1}{2} u\hook\u) . 
\end{equation}

Then the equations for $S$, $\rho$, $\u$, and $\w$ 
have the elegant transport formulation 
\begin{align}
& \Dt S =0 , 
\label{entrflow}\\
& \Dt\rho = -\rho \div_g \u , 
\label{densflow}\\
& \Dt\u = \d\big(\tfr{1}{2}|\u|_g^2 -e-\rho^{-1}p\big) +e_S \d S , 
\quad
d\u =\w , 
\label{velflow}\\
& \Dt\w = \d e_S\wedge \d S , 
\quad
d\w =0 , 
\label{curlflow}
\end{align}
with $\div_g \u =\gnabla\hook\u=\div\, u$ and $|\u|_g^2=u\hook\u=g(u,u)$,
where 
\begin{equation}
e=\textstyle\int\rho^{-2}p d\rho
\end{equation} 
is the thermodynamic energy of the fluid.
In particular, note 
$e=\int\rho^{-2}P(\rho,S) d\rho$ is a specified function of $\rho$ and $S$
for compressible flow, 
whereas $e=0$ (since $\d\rho=0$) vanishes for incompressible flow.

\section{Moving surfaces and conserved integrals}

Now consider any smooth orientable $p$-dimensional submanifold $\S(t)\subset M$
that is transported along streamlines in the fluid. 
In particular, in local coordinates, 
each point $x^i \in \S(t)$ obeys $dx^i/dt= \Lu x^i= u\hook\nabla x^i$ 
($i=1,\ldots,n$). 
A \emph{conserved integral} on the moving surface $\S(t)$ consists of 
an integral continuity equation given by 
\begin{equation}
\fr{d}{dt}\int_{\S(t)} \mbf\alpha 
= \int_{\p\S(t)} \mbf\beta
\label{Scontinuity}
\end{equation}
for a $p$-form density $\mbf\alpha$ and a $p-1$-form flux $\mbf\beta$ 
that are functions of the space and time coordinates $x^i$, $t$, 
and the fluid variables $S$, $\rho$, $\u$, $\w$ 
(and possibly their spatial derivatives) 
subject to the fluid equations \eqref{entrflow}--\eqref{velflow}. 

Integrals of this form \eqref{Scontinuity} will define a \emph{\com/}
on the moving surface $\S(t)$ 
for all formal solutions of the fluid equations
provided that the flux integral on $\p\S(t)$ vanishes identically. 
If $\S(t)$ is boundaryless 
then every conserved integral \eqref{Scontinuity} yields a \com/. 
Alternatively, if $\S(t)$ has a boundary 
then a conserved integral \eqref{Scontinuity} yields a \com/ 
only when $\mbf\beta$ is an exact $p-1$ form. 
Note that $\S(t)$ will be a moving domain $\V(t)$ if $p=n$,
in which case the conserved integral \eqref{Scontinuity} coincides with
the continuity equation
\begin{equation}
\fr{d}{dt}\int_{\V(t)} T dV
= -\int_{\p\V(t)} g(X,\nor) dA
\label{Vcontinuity}
\end{equation}
given by 
\begin{equation}
\mbf\alpha = \mbf\epsilon T ,
\quad 
\mbf\beta = -X\hook\mbf\epsilon
\label{TX}
\end{equation}
where $dV=\mbf\epsilon_g$ is the volume $n$-form 
dual to the tensor $\epsilon_g$,
and $dA=\nor\hook\mbf\epsilon_g$ is the hypersurface area $n-1$-form
in terms of the unit-normal vector $\nor$. 
In this situation the conservation law \eqref{conslaw} reduces to 
a space-time divergence $D_t T + \Div X=0$ 
where $\Div$ denotes $\div$ acting as a total derivative. 

\begin{proposition}
A $p$-form density $\mbf\alpha$ and a $p-1$-form flux $\mbf\beta$ 
yield a conserved integral \eqref{Scontinuity} 
on all $p$-dimensional moving surfaces $S(t)\subset M$ 
iff 
\begin{equation}
\Dt\mbf\alpha = \d\mbf\beta
\label{conslaw}
\end{equation}
holds for all formal solutions of the fluid equations. 
\end{proposition}

Proof: 
Let $\phi_t$ be the diffeomorphism of $M$ defined by 
the streamlines of the vector field $u\big|_t$ with $\phi_0=\id$. 
The moving surface $S(t)$ can then be viewed as the flow of a fixed surface
$S(0)$ in $M$. 
Under this flow, the pullback of $\mbf\alpha\big|_{S(t)}$ is given by 
$\phi^*_t\mbf\alpha\big|_{S(0)}$, 
where 
$\dfrac{d}{dt}(\phi^*_t\mbf\alpha) = D_t\mbf\alpha +\Lu\mbf\alpha=\Dt\mbf\alpha$.
Hence we have
\begin{equation}
\bigg(\fr{d}{dt}\int_{\S(t)} \mbf\alpha \bigg)\bigg|_{t=0}
= \int_{\S(0)} \fr{d}{dt}(\phi^*_t\mbf\alpha) 
= \int_{\S(0)} \Dt\mbf\alpha
\end{equation}
and thus the integral continuity equation \eqref{Scontinuity} becomes
\begin{equation}
0 = \int_{\S(0)} \Dt\mbf\alpha - \int_{\p\S(0)} \mbf\beta
= \int_{\S(0)} (\Dt\mbf\alpha - \d\mbf\beta) . 
\end{equation}
The vanishing of this integral for an arbitrary $p$-dimensional surface $S(0)$ 
is equivalent to the conservation law \eqref{conslaw}. $\blacksquare$

Conserved vorticity integrals 
in which the density $\mbf\alpha$ and the flux $\mbf\beta$ 
have an essential dependence on the curl $2$-form $\w$ will now be derived 
from the transport equations \eqref{entrflow}--\eqref{curlflow}, 
first for isentropic fluid flow and then for non-isentropic fluid flow.

\section{Helicity on moving surfaces}

Three-dimensional helicity \eqref{3dhelicity} can be naturally generalized
to all odd dimensions $n>3$ in terms of the vorticity vector 
\cite{AncDar1,ArnKhe}
\begin{equation}
\vortvec = *(\underbrace{\w\wedge\cdots\wedge\w}_{(n-1)/2 \text{ times}})
= \epsilon_g\hook \w^{(n-1)/2}
\end{equation}
defined on the $n$-dimensional Riemannian manifold $M$.
Here $*$ denotes the Hodge dual operator, 
acting by contraction with respect to the volume tensor $\epsilon_g$,
and powers of $\w$ denote exterior products. 
For a moving domain $\V(t)\subset M$, 
the $n$-dimensional helicity integral for isentropic fluid flow in $M$ 
is given by a continuity equation \eqref{Vcontinuity} 
holding for the conserved density $T=\vortvec\hook\u$ 
and the spatial flux $X = -\sigma\vortvec$
with $\sigma= \tfrac{1}{2} |\u|_g^2 -e -p/\rho$,
where $e$ is the thermodynamic energy of the fluid. 
A simpler formulation is obtained by using the corresponding 
differential forms \eqref{TX}:
$\mbf\alpha = \u\wedge \w^{(n-1)/2}$ and $\mbf\beta = \sigma\w^{(n-1)/2}$.
The resulting conserved helicity integral \eqref{Scontinuity} 
has a natural extension from moving domains in $M$ 
to odd-dimensional moving surfaces in $M$, 
whether $n$ is odd or even.

\begin{theorem}
For isentropic inviscid fluid flow in any Riemannian manifold $M$
with dimension $n\ge 1$, 
the $2q+1$-form density $\mbf\alpha = \u\wedge\w^q$ 
and the $2q$-form flux $\mbf\beta = (\tfrac{1}{2}|\u|_g^2 +e -p/\rho) \w^q$ 
satisfy the continuity equation
\begin{equation}
\fr{d}{dt}\int_{\S(t)} \u\wedge \w^q 
= \int_{\p\S(t)} (\tfrac{1}{2}|\u|_g^2 +e -p/\rho) \w^q ,\quad
0\le q\le (n-1)/2
\label{helicityconslaw}
\end{equation}
yielding a conserved helicity integral 
on any smooth orientable submanifold $\S(t)$,
with odd dimension $2q+1$, transported along streamlines in the fluid. 
\end{theorem}
 
Proof of conservation: 
Isentropic fluid flow has $\d S=0$ in $M$, 
so thus the velocity and curl equations \eqref{velflow}--\eqref{curlflow}
reduce to 
\begin{equation}
\Dt\w = 0 ,\quad
\Dt\u = \d\sigma ,\quad
\sigma = \tfrac{1}{2}|\u|_g^2 -e-\rho^{-1}p
\label{isentropicfluideqns}
\end{equation}
with 
\begin{equation}
\d\w =0 ,\quad
\d\u =\w .
\label{dcurlvel}
\end{equation}
Since $\Dt\u$ is an exact $1$-form 
and $\w$ is a closed $2$-form annihilated by $\Dt$,
we have 
$\Dt( \u\wedge \w^q ) =\d\sigma\wedge \w^q = \d(\sigma\w^q)$,
which establishes \eqref{helicityconslaw}. $\blacksquare$

When the surface $\S$ is a closed curve $\C$, 
the helicity integral \eqref{helicityconslaw} 
(with $q=0$ and $\p\S=0$)
reduces to the circulation integral
\begin{equation}
\fr{d}{dt}\oint_{\C(t)} \u =0 
\label{circcom}
\end{equation}
providing a generalization of Kelvin's circulation theorem 
to isentropic fluid flow (compressible or incompressible) 
in Riemannian manifolds of any dimension $n>1$.

\section{Enstrophy on moving surfaces}

Two-dimensional enstrophy \eqref{2denstrophy} has a natural extension 
to all even dimensions $n>2$ in terms of the vorticity scalar
\cite{AncDar1,ArnKhe}
\begin{equation}
\vortscal = *(\underbrace{\w\wedge\cdots\wedge\w}_{n/2 \text{ times}})
= \epsilon_g\hook \w^{n/2}
\label{Mvorticity}
\end{equation}
defined on the $n$-dimensional Riemannian manifold $M$ 
in terms of the Hodge dual operator $*$. 
Here powers of $\w$ again denote exterior products. 
For a moving domain $\V(t)\subset M$ in isentropic fluid flow, 
the $n$-dimensional enstrophy integral 
is given by a continuity equation \eqref{Vcontinuity} 
holding for the conserved density $T=f(\vortscal/\rho) \vortscal$
with no spatial flux $X =0$,
where $f$ is an arbitrary non-constant function. 

An equivalent formulation uses the corresponding differential forms \eqref{TX}:
$\mbf\alpha = f(\vortscal/\rho)\w^{n/2}$ and $\mbf\beta = 0$. 
The resulting conserved enstrophy integral \eqref{Scontinuity} 
can be generalized from moving domains in $M$ 
to even-dimensional moving surfaces in $M$. 

\begin{theorem}
For isentropic inviscid fluid flow in any Riemannian manifold $M$
with even dimension $n\ge 2$, 
the $2q$-form density $\mbf\alpha = f\w^q$ 
depending on an arbitrary non-constant function $f(\vortscal/\rho)$
satisfies the continuity equation
\begin{equation}
\fr{d}{dt}\int_{\S(t)} f\w^q = 0 ,\quad
1\le q\le n/2
\label{enstrophyconslaw}
\end{equation}
yielding a conserved enstrophy integral 
on any smooth orientable submanifold $\S(t)$,
with even dimension $2q$, transported along streamlines in the fluid. 
\end{theorem}
 
Proof of conservation: 
The curl equation \eqref{isentropicfluideqns} 
holding for isentropic fluid flow implies that 
$\Dt(f\w^q) = \Dt(\vortscal/\rho)f'\w^q$,
where $\vortscal = \epsilon_g\hook\w^{n/2}$. 
Then using 
\begin{equation}
\Dt\vortscal = (\Dt\epsilon_g) \hook\w^{n/2}, 
\quad
\Dt\epsilon_g = \Lu\epsilon_g =-(\div\, u)\epsilon_g ,
\quad
\Dt\rho = -(\div\, u) \rho , 
\end{equation}
we have 
\begin{equation}
\Dt(\rho^{-1}\epsilon_g) = 0
\label{invariant}
\end{equation}
whence $\Dt(\vortscal/\rho) = 0$. 
This implies $\Dt(f\w^q) =0$, 
establishing \eqref{enstrophyconslaw}. $\blacksquare$

The enstrophy integral \eqref{enstrophyconslaw} 
can be alternatively expressed as 
\begin{equation}
\fr{d}{dt}\int_{\S(t)} f\w^q 
= -\fr{d}{dt}\int_{\S(t)} f' \d(\vortscal/\rho)\wedge\u\wedge\w^{q-1} 
+ \fr{d}{dt}\oint_{\p\S(t)} f\u\wedge\w^{q-1} 
=0
\label{enstrophyconslaw2}
\end{equation}
through the properties \eqref{dcurlvel} of $\w$,
whether the surface $\S$ has a boundary or is boundaryless. 
Consequently, if the function $f$ is constant,
then the enstrophy will reduce to the helicity on the boundary of $\S$
or will otherwise vanish when $\S$ has no boundary.

\section{Entropy circulation on moving surfaces}

Similarly to enstrophy and helicity, 
two-dimensional entropy circulation \eqref{2dentropycirc}
and three-dimensional entropy circulation \eqref{3dentropycirc} 
have a straightforward extension respectively to 
all even dimensions $n>2$ and all odd dimensions $n>3$. 
The entropy circulation integral 
for a moving domain $\V(t)\subset M$ in non-isentropic fluid flow 
is given by a continuity equation \eqref{Vcontinuity} 
that holds in the even-dimensional case 
for the conserved density $T=f(S)\vortscal$ 
and the spatial flux $X =-f(S) e_S (\p\vortscal/\p\w)\hook \d S$,
where $e$ is the thermodynamic energy of the fluid,
and that holds in the odd-dimensional case 
for the conserved density $T=f(\circ/\rho)\circ$ 
with vanishing spatial flux $X =0$,
where 
\begin{equation}
\circ= *(\d S\wedge\w^{(n-1)/2}) = \vortvec\hook \d S
\label{Mentrcirc}
\end{equation}
defines the entropy circulation scalar on the Riemannian manifold $M$
in terms of the vorticity vector $\vortvec$. 
The corresponding differential forms \eqref{TX} are given by 
$\mbf\alpha = f(S)\w^{n/2}$ 
and $\mbf\beta = \tfrac{1}{2}n f(S) e_S \d S\wedge\w^{(n-2)/2}$ 
when the dimension $n$ is even, 
and by $\mbf\alpha = f(\circ/\rho)\w^{(n-1)/2}$ and $\mbf\beta = 0$
when the dimension $n$ is odd. 
In each case, there is a natural generalization of the resulting 
conserved entropy circulation integral \eqref{Scontinuity} 
to moving surfaces in $M$. 

The generalization is presented first for even-dimensional surfaces. 

\begin{theorem}
For non-isentropic inviscid fluid flow in any Riemannian manifold $M$
with even or odd dimension $n\ge 2$, 
the $2q$-form density $\mbf\alpha = f\w^q$ 
and the $2q-1$-form flux $\mbf\beta = q e_S f \d S \wedge \w^{q-1}$ 
depending on an arbitrary non-constant function $f(S)$
satisfy the continuity equation
\begin{equation}
\fr{d}{dt}\int_{\S(t)} f\w^q = 
q \oint_{\p\S(t)} e_S f \d S\wedge \w^{q-1} ,\quad
1\le q\le n/2
\label{entropycircconslawSdimeven}
\end{equation}
yielding a conserved entropy circulation integral 
on any smooth orientable submanifold $\S(t)$,
with even dimension $2q$, transported along streamlines in the fluid. 
\end{theorem}
 
Proof of conservation: 
From the entropy equation \eqref{entrflow}
and the curl equation \eqref{curlflow}
holding for non-isentropic fluid flow, 
we have 
$\Dt(f\w^q) = q f \d e_S\wedge \d S\wedge \w^{q-1}
= q \d( e_S f \d S\wedge \w^{q-1} )$
since the $2$-form $\w$ is closed. 
This establishes \eqref{entropycircconslawSdimeven}. $\blacksquare$

When the surface $\S$ is boundaryless, 
the entropy circulation integral \eqref{entropycircconslawSdimeven} 
becomes 
\begin{equation}
\fr{d}{dt}\oint_{\S(t)} f\w^q 
= -\fr{d}{dt}\oint_{\S(t)} f' \d S\wedge\u\wedge\w^{q-1} 
=0
\label{entropycirccom}
\end{equation}
from the properties \eqref{dcurlvel} of $\w$. 
As a consequence, if the gradient of the entropy $S$ is aligned 
with the fluid streamlines on the surface $\S$
(which will occur whenever the fluid flow is isentropic in $\S\subset M$), 
then the entropy circulation \eqref{entropycirccom} 
will vanish due to $\d S\wedge\u=0$.

The analogous generalization for odd-dimensional surfaces is presented next. 

\begin{theorem}
For non-isentropic inviscid fluid flow in any Riemannian manifold $M$
with odd dimension $n\ge 3$, 
the $2q+1$-form density $\mbf\alpha = f \d S\wedge\w^q$ 
depending on an arbitrary non-constant function $f(\circ/\rho)$
satisfies the continuity equation
\begin{equation}
\fr{d}{dt}\int_{\S(t)} f \d S\wedge\w^q = 0 ,\quad
1\le q\le (n-1)/2
\label{entropycircconslawSdimodd}
\end{equation}
yielding a conserved entropy circulation integral 
on any smooth orientable submanifold $\S(t)$,
with odd dimension $2q+1$, transported along streamlines in the fluid. 
This integral remains conserved if $f$ is generalized to be a function of 
both $\circ/\rho$ and $S$. 
\end{theorem}

Proof of conservation: 
The curl equation \eqref{curlflow} 
combined with the entropy equation \eqref{entrflow} yield
$\Dt(\d S\wedge\w^q) = 0$ 
and 
$\Dt f= f_{\tilde\circ} \Dt\tilde\circ 
= f_{\tilde \circ} \Dt(\rho^{-1}\epsilon_g)\hook (\d S\wedge \w^{(n-1)/2})$ 
with $\tilde\circ =\circ/\rho$ 
where $\circ=\epsilon_g\hook (\d S\wedge \w^{(n-1)/2})$. 
Then equation \eqref{invariant} implies 
$\Dt f=0$ and hence $\Dt(f \d S\wedge\w^q) =0$,
which establishes \eqref{entropycircconslawSdimodd}. $\blacksquare$

In contrast to the even-dimensional integral \eqref{entropycircconslawSdimeven},
here the odd-dimensional entropy circulation integral \eqref{entropycircconslawSdimodd} 
can be expressed as 
\begin{equation}
\fr{d}{dt}\int_{\S(t)} f \d S\wedge\w^q 
= -\fr{d}{dt}\int_{\S(t)} f' S \d(\circ/\rho)\wedge\w^q 
+ \fr{d}{dt}\oint_{\p\S(t)} f S\w^q
=0
\label{entropycirccom2}
\end{equation}
whether the surface $\S$ has a boundary or is boundaryless. 
If the function $f$ is constant,
then this integral \eqref{entropycirccom2} 
will reduce to the entropy circulation on the boundary of $\S$
or will otherwise vanish when $\S$ has no boundary.

\section{New constants of motion}

As a preliminary step, 
for any $s$-dimensional smooth orientable surface $\S\subset M$, 
let $\{\e{\mu}\}_{\mu=1,\ldots,n-s}$ be an arbitrary orthonormal basis 
for the normal space of $\S$ in $T_x M$ at each point of $\S$. 
Then the volume forms of the surface $\S$ and its boundary $\p\S$ 
can be defined by the projections 
\begin{equation}
dV_{\S}=\mbf\epsilon(\S) 
= (\e{1}\wedge \cdots\wedge \e{n-s})\hook \mbf\epsilon_g 
\end{equation}
and 
\begin{equation}
dA_{\S}=\nor_{\S}\hook\mbf\epsilon(\S) =\mbf\epsilon(\p\S) 
= (\e{1}\wedge \cdots\wedge \e{n-s}\wedge \nor_{\S})\hook \mbf\epsilon_g
\end{equation}
where $\nor_{\S}$ denotes the unit normal of $\p\S$ in $\S$
and $\mbf\epsilon_g$ is the volume $n$-form of $M$. 
Additionally, the dual volume tensors 
$\epsilon(\S)$ for $\S$ 
and $\epsilon(\p\S)$ for $\p\S$ 
can be defined in terms of the volume tensor $\epsilon_g$ for $M$ by
\begin{equation}\label{Svoltens}
\e{1}\wedge \cdots\wedge \e{n-s} \wedge \epsilon(\S)= \epsilon_g ,
\quad
\nor_{\S} \wedge \epsilon(\p\S)= \epsilon(\S) . 
\end{equation}

Now, if $\S(t)$ is an odd-dimensional surface 
transported along fluid streamlines in $M$, 
the helicity integral \eqref{helicityconslaw} 
for isentropic fluid flow 
and the entropy circulation integral \eqref{entropycircconslawSdimodd}
for non-isentropic fluid flow 
can be formulated respectively as 
\begin{equation}
\fr{d}{dt}\int_{\S(t)} \vortvec_{\S}\hook \u_{\S} \;dV_{\S}
= \oint_{\p\S(t)} (\tfrac{1}{2} |\u|_g^2 -e -p/\rho) g(\vortvec_{\S},\nor_{\S}) \;dA_{\S}
\label{helicity}
\end{equation}
and
\begin{equation}
\fr{d}{dt}\int_{\S(t)} f(S,\circ/\rho) \vortvec_{\S}\hook\d_{\S} S \;dV_{\S}
= 0 . 
\label{entrcircSodd}
\end{equation}
Here 
\begin{equation}
\vortvec_{\S} = 
\epsilon(\S)\hook(\w|_{\S})^q
\label{vorticityvec}
\end{equation}
defines the vorticity vector of the surface $\S$ with $q=\tfrac{1}{2}(\dim(\S)-1)$,
while $\u_{\S}$ and $\d_{\S}S$ denote the respective projections of 
the velocity $1$-form $\u$ and the gradient $1$-form $\d S$
into the cotangent space of $\S$, 
and $\circ$ is the entropy circulation scalar \eqref{Mentrcirc} of the manifold $M$ when the dimension $n$ is odd. 

Similarly, if $\S(t)$ is an even-dimensional surface 
transported along fluid streamlines in $M$, 
the enstrophy integral \eqref{enstrophyconslaw} 
for isentropic fluid flow 
and the entropy circulation integral \eqref{entropycircconslawSdimeven} 
for non-isentropic fluid flow 
have the respective formulations
\begin{equation}
\fr{d}{dt}\int_{\S(t)} f(\vortscal/\rho) \vortscal_{\S} \;dV_{\S}
= 0 
\label{enstrophy}
\end{equation}
and 
\begin{equation}
\fr{d}{dt}\int_{\S(t)} f(S) \vortscal_{\S} \;dV_{\S}
= q\oint_{\p\S(t)} e_S f(S) g(\vortvec_{\p\S}, \nabla_{\p\S} S) \;dA_{\S}
\label{entrcircSeven}
\end{equation}
where 
\begin{equation}
\vortscal_{\S} = 
\epsilon(\S)\hook(\w|_{\S})^q
\label{vorticityscal}
\end{equation}
defines the vorticity scalar of the surface $\S$ with $q=\tfrac{1}{2}\dim(\S)$,
while $\nabla_{\p\S}$ denotes the projection of the gradient $\gnabla$ 
into the tangent space of $\p\S$, 
and $\vortscal$ is the vorticity scalar \eqref{Mvorticity} of the manifold $M$ 
when the dimension $n$ is even. 

These formulations \eqref{helicity}--\eqref{vorticityscal} are useful 
for investigating the \csom/ that arise under various conditions
for moving surfaces $\S(t)$ in isentropic and non-isentropic fluid flow in $M$ 
as follows. 

\subsection{Closed moving surfaces spanned by a moving hypersurface}

Consider any even-dimensional surface $\Sigma(t)$ spanning 
a closed orientable surface $\S(t)$ 
transported along the fluid streamlines in $M$,
with $\dim\S=\dim\Sigma-1$, $\p\Sigma=\S$ and $\p\S=0$. 
Then Stokes' theorem can be applied to convert 
the conserved helicity integral \eqref{helicity} 
into the equivalent form 
\begin{equation}
\fr{d}{dt}\oint_{\S(t)} \vortvec_{\S}\hook\u_{\S} \;dV_{\S}
= \fr{d}{dt}\int_{\Sigma(t)} \vortscal_{\Sigma} \;dV_{\Sigma}
=0
\label{evendimvorticitycom}
\end{equation}
yielding a scalar vorticity \com/ on $\Sigma(t)$
for $n$-dimensional isentropic (compressible or incompressible) fluid flow. 
This new \com/ \eqref{evendimvorticitycom} 
provides a higher-dimensional version of Kelvin's circulation \eqref{circcom}, 
which has the analogous formulation as a vorticity \com/ 
\begin{equation}
\fr{d}{dt}\oint_{\C} u_{\S}\; ds 
= \fr{d}{dt}\int_{\Sigma(t)} \vortscal_{\Sigma} \;dV_{\Sigma}
=0
\label{curvevorticitycom}
\end{equation}
on $2$-surfaces $\Sigma(t)$ that span a closed curve $\C(t)$ 
transported along the fluid streamlines
in any even or odd dimension $n\geq 2$. 
(Note here $u_{\S}=\epsilon(\C)\hook\u$ denotes the scalar projection of $\u$
along the curve $\C$,
and $ds=\mbf\epsilon(\C)$ is the arclength $1$-form of $\C$.)
When the dimension $n$ is even, 
these \csom/ \eqref{evendimvorticitycom} and \eqref{curvevorticitycom}
are special cases of the enstrophy \com/ \eqref{enstrophy} on $\Sigma(t)$. 
Therefore, 
the helicity and enstrophy integrals can be unified in the form of 
a generalized scalar vorticity \com/ 
\begin{equation}
\fr{d}{dt}\int_{\Sigma(t)} f(\vortscal/\rho) \vortscal_{\Sigma} \;dV_{\Sigma}
= 0 
\label{generalvorticitycom}
\end{equation}
in which $f$ is a function of $\vortscal/\rho$ when $n$ is even 
or $f$ is a constant when $n$ is odd. 

In contrast, 
for $n$-dimensional non-isentropic (compressible) fluid flow, 
the scalar vorticity integral \eqref{generalvorticitycom} 
is no longer conserved but instead is replaced by 
the conserved entropy circulation integrals 
\eqref{entrcircSodd} and \eqref{entrcircSeven}. 
Given any odd-dimensional surface $\Sigma(t)$ spanning 
a closed orientable surface $\S(t)$ 
transported along the fluid streamlines in $M$, 
with $\dim\S=\dim\Sigma-1$, $\p\Sigma=\S$ and $\p\S=0$, 
the integral \eqref{entrcircSeven} can be converted through Stokes' theorem
into a special case of the integral \eqref{entrcircSodd} on $\Sigma(t)$ 
when the dimension $n$ is odd,
\begin{equation}
\fr{d}{dt}\oint_{\S(t)} f(S) \vortscal_{\S} \;dV_{\S}
= \fr{d}{dt}\int_{\Sigma(t)} f'(S) \vortvec_{\Sigma}\hook\d S \;dV_{\Sigma}
=0 . 
\end{equation}
As a result, 
both entropy circulation integrals \eqref{entrcircSodd} and \eqref{entrcircSeven} 
have a unified formulation 
\begin{equation}
\fr{d}{dt}\int_{\Sigma(t)} f(S,\circ/\rho) 
\vortvec_{\Sigma}\hook \d S \;dV_{\Sigma} 
= 0
\label{nonisenvorticitycom}
\end{equation}
yielding a new vorticity \com/ on $\Sigma(t)$
for $n$-dimensional non-isentropic (compressible) fluid flow. 
This \com/ \eqref{nonisenvorticitycom} measures the alignment between 
the entropy gradient and the vorticity vector of the fluid on $\Sigma(t)$,
weighted by a function $f$ of $S$ when $n$ is even 
or a function $f$ of both $S$ and $\circ/\rho$ when $n$ is odd. 

\subsection{Non-closed moving surfaces and vorticity boundary conditions}

Consider an orientable non-closed, but otherwise arbitrary, surface $\S(t)$ 
transported along the fluid streamlines in $M$. 
Then the entropy circulation integral \eqref{entrcircSodd} 
when $\S$ is odd-dimensional 
and the enstrophy integral \eqref{enstrophy} 
when $\S$ is even-dimensional 
are \csom/ respectively for non-isentropic fluid flow and isentropic fluid flow 
in $M$. 
In contrast, 
the helicity integral \eqref{helicity} when $\S$ is odd-dimensional 
and the entropy circulation integral \eqref{entrcircSeven} when $\S$ is even-dimensional 
are \csom/ if and only if the flux integrals on the boundary $\p\S(t)$
vanish respectively for isentropic fluid flow and non-isentropic fluid flow in $M$. 

The helicity flux integral \eqref{helicity} clearly vanishes if 
$g(\vortvec_{\S},\nor_{\S})|_{\p\S}=0$ holds on the boundary surface $\p\S(t)$. 
This condition states that the vorticity vector $\vortvec_{\S}$ at $\p\S(t)$
must lie in the tangent space of $\p\S(t)$. 
A useful equivalent formulation is that 
the vorticity scalar of the boundary surface $\p\S(t)$ must vanish
everywhere on $\p\S(t)$, 
\begin{equation}
\vortscal_{\p\S} =0 
\label{helicityflux}
\end{equation}
since $g(\vortvec_{\S},\nor_{\S})|_{\p\S}=\vortscal_{\p\S}$ holds 
due to the identity \eqref{Svoltens} relating the volume tensors of 
$\S(t)$ and $\p\S(t)$. 

Likewise, the entropy circulation flux integral \eqref{entrcircSeven} 
vanishes whenever 
$g(\vortvec_{\p\S}, \nabla_{\p\S} S)|_{\p\S}=0$,
so that the vorticity vector $\vortvec_{\p\S}$ 
must be orthogonal to the entropy gradient at the boundary surface $\p\S(t)$.
This condition has the equivalent formulation
\begin{equation}
\vortvec_{\p\S}\hook\d_{\p\S} S=0
\label{entropycircflux}
\end{equation}
where $\d_{\p\S}$ denotes the projection of the exterior derivative 
into the cotangent space of $\p\S$. 

As a consequence of the following result, 
these two flux conditions \eqref{helicityflux} and \eqref{entropycircflux}
are readily seen to be preserved under transport along fluid streamlines. 

\begin{lemma}
For isentropic fluid flow in $M$, 
\begin{equation}
\Dt\vortscal_{\p\S} = -\vortscal_{\p\S} \div_{\p\S} u
, 
\end{equation}
and for non-isentropic fluid flow in $M$, 
\begin{equation}
\Dt(\vortvec_{\p\S}\hook\d_{\p\S} S)
= -(\vortvec_{\p\S}\hook\d_{\p\S} S) \div_{\p\S} u ,
\end{equation}
where $\div_{\p\S}$ denotes the divergence operator projected into the boundary surface $\p\S$. 
\end{lemma}

Proof:
View $\S(t)$ as the flow of a fixed surface $\S(0)$
under the diffeomorphism $\phi_t$ of $M$ generated by the streamlines $u\big|_t$
with $\phi_0=\id$. 
The pullback of the boundary volume form 
$dA_{\S}=\mbf\epsilon(\p\S(t))$ of $\p\S(t)$ 
yields
$\dfrac{d}{dt}\phi^*_t dA_{\S}\big|_{t=0} = \Lu dA_{\S}\big|_{t=0} 
= (\div_{\p\S} u) dA_{\S}$ on the boundary surface $\p\S(0)$. 
This implies 
$\Lu\mbf\epsilon(\p\S) = (\div_{\p\S}u)\mbf\epsilon(\p\S)$ on $\p\S$, 
and hence the dual volume tensor $\epsilon(\p\S)$ obeys
$\Lu\epsilon(\p\S) = -(\div_{\p\S}u)\epsilon(\p\S)$. 
Now let $q=\dim\p\S$. 
From 
$\vortscal_{\p\S} = \epsilon(\p\S)\hook\w^{q/2}$,
we have 
$\Dt\vortscal_{\p\S} = \Lu\epsilon(\p\S)\hook\w^{q/2}
= -(\div_{\p\S}u)\epsilon(\p\S)\hook\w^{q/2}$
since $\Dt\w=0$ holds for 
isentropic (compressible or incompressible) fluid flow. 
Similarly, from 
$\vortvec_{\p\S}\hook\d_{\p\S} S = \epsilon(\p\S)\hook(\d S\wedge\w^{(q-1)/2})$,
we get 
$\Dt(\vortvec_{\p\S}\hook\d_{\p\S} S) 
= \Lu\epsilon(\p\S)\hook(\d S\wedge\w^{(q-1)/2})
= -(\div_{\p\S}u)\epsilon(\p\S)\hook(\d S\wedge\w^{(q-1)/2})$
since $\Dt(\w\wedge\d S)=0$ holds for non-isentropic fluid flow. $\blacksquare$

Therefore, 
when an orientable non-closed odd-dimensional surface $\S(t)$ satisfies 
the vorticity boundary condition \eqref{helicityflux} 
transported along fluid streamlines in $M$, 
the helicity integral \eqref{helicity} yields a \com/
\begin{equation}
\fr{d}{dt}\int_{\S(t)} \vortvec_{\S}\hook \u_{\S} \;dV_{\S}
= 0
\label{helicitySnonclosed}
\end{equation}
for isentropic fluid flow in $M$. 
Similarly, 
when an orientable non-closed even-dimensional surface $\S(t)$ satisfies 
the vorticity boundary condition \eqref{entropycircflux}
transported along fluid streamlines in $M$, 
the entropy circulation integral \eqref{entrcircSeven} yields a \com/
\begin{equation}
\fr{d}{dt}\int_{\S(t)} f(S) \vortscal_{\S} \;dV_{\S}
= 0 
\label{entrcircSevennonclosed}
\end{equation}
for non-isentropic fluid flow in $M$.

\section{Concluding remarks}

There are several interesting directions for future work. 

First, 
the physical meanings of helicity, enstrophy, circulation and entropy circulation
are fairly well understood when $M=\Rnum^3,\Rnum^2$. 
In higher dimensions, 
what is the precise physical content of the new \csom/?
This could be elucidated by evaluating the integrals 
\eqref{evendimvorticitycom} and \eqref{nonisenvorticitycom} in $M=\Rnum^n$
for some physically relevant fluid configurations 
as well as for analytically interesting exact solutions of the fluid equations. 

Second, 
the helicity integral for a moving domain $\V(t)$ in $M=\Rnum^3$ 
is well-known to equal the average linking number of the integral curves of
the vorticity vector $\vortvec$ \cite{ArnKhe}. 
What is the relation between 
the new helicity/circulation \csom/ 
\eqref{evendimvorticitycom} and \eqref{helicitySnonclosed}
for moving surfaces in $M=\Rnum^n$ with $n>3$
and the topological linking of vorticity-lines? 
Likewise, is there an interpretation of the new entropy circulation \csom/ 
\eqref{nonisenvorticitycom} and \eqref{entrcircSevennonclosed}
in terms of the topological structure of 
vorticity-lines and entropy gradient-lines
for moving surfaces in $M=\Rnum^n$? 

Third, 
what information do these new \csom/ contain when the manifold $M$ 
or the moving surface have nontrivial homology? 

Finally, another interesting open question is whether the fluid equations 
\eqref{veleqn}--\eqref{curleqn} 
admit any additional conserved integrals that yield \csom/ 
on moving surfaces in $n>1$ dimensions, 
other than the conserved mass integral 
\begin{equation}
\fr{d}{dt}\int_{\V(t)} \rho dV
= 0
\label{masscom}
\end{equation}
for $n$-dimensional domains $\V(t)$ transported along fluid streamlines 
in a Riemannian manifold $M$.

\appendix
\section{Coordinate formulation}

It is worthwhile to write the new conserved integrals 
\eqref{helicityconslaw}, \eqref{enstrophyconslaw}, 
\eqref{entropycircconslawSdimeven} and \eqref{entropycircconslawSdimodd}
in terms of the physical fluid variables 
$u^i$ and $\omega^{ij}=2\nabla^{[i}u^{j]}$
given by local coordinates $x^i$ on $M$. 
Note the coordinate formulation of the transport equations 
\eqref{velflow} and \eqref{curlflow} for these variables is given by 
\begin{align}
& u^i_t+u^j\nabla_j u^i = -\rho^{-1}\nabla^i p 
\label{coordveleqn}\\
& \omega^{ij}_t +u^k\nabla_k \omega^{ij} -2\omega_k{}^{[i} \sigma^{j]k} 
= \rho^{-2}\nabla^{[i}\rho \nabla^{j]} p , \quad
\nabla^{[k} \omega^{ij]}=0 
\end{align}
where $\sigma^{ij}=2\nabla^{(i}u^{j)}$ is the symmetric derivative of $u^i$. 

Now consider isentropic fluid flow, \ie/ $S=\const$ in $M$, 
with 
\begin{equation}
p=P(\rho) ,\quad
\rho_t + \nabla_i(\rho u^i) = 0 
\label{coorddenspreseqns}
\end{equation}
when the fluid is compressible, 
or 
\begin{equation}
\rho=\const ,\quad
-\rho^{-1} \nabla^i\nabla_i p = \nabla_i u^j \nabla_j u^i + R_{ij} u^i u^ j
\label{coordpreseqn}
\end{equation}
when the fluid is incompressible,
where $R_{ij}$ is the Ricci tensor 
and $\nabla_i u^j \nabla_j u^i = \tfrac{1}{2}( \sigma_{ij}\sigma^{ij} - \omega_{ij}\omega^{ij})$ is the difference of the norms of 
the symmetric derivative $\sigma^{ij}=2\nabla^{(i}u^{j)}$ 
and the curl $\omega^{ij}$ of $u^i$. 
Then the helicity integral \eqref{helicityconslaw} 
for any orientable odd-dimensional surface $\S(t)$ 
transported along fluid streamlines in $M$ 
is given by 
\begin{equation}
\fr{d}{dt}\int_{\S(t)} g_{ij} u_{\S}^i \vortvec_{\S}^j \;dV_{\S}
= \int_{\p\S(t)} (\tfrac{1}{2} g_{kl} u^k u^l -e -p/\rho) g_{ij}\vortvec_{\S}^i \nor_{\S}^j \;dA_{\S}
\end{equation}
in terms of the vorticity vector $\vortvec_{\S}^i$ of the surface $\S$
and the projection $u_{\S}^i$ of the velocity vector $u^i$ 
into the tangent space of $\S$.
Similarly, 
for any orientable even-dimensional surface $\S(t)$ 
transported along fluid streamlines in $M$, 
the enstrophy integral \eqref{enstrophyconslaw} is given by 
\begin{equation}
\fr{d}{dt}\int_{\S(t)} f(\vortscal/\rho) \vortscal_{\S} \;dV_{\S}
= 0 
\end{equation}
in terms of the vorticity scalar $\vortscal_{\S}$ of the surface $\S$, 
where $\vortscal$ is the vorticity scalar \eqref{Mvorticity} of the manifold $M$
when the dimension $n$ is even. 

Next consider non-isentropic fluid flow, \ie/ $S\neq \const$ in $M$, 
with 
\begin{equation}
p=P(\rho,S) ,\quad
\rho_t + \nabla_i(\rho u^i) = 0 ,\quad
S_t + u^i\nabla_i S =0 . 
\label{coorddensentreqns}
\end{equation}
Then for any orientable odd-dimensional surface $\S(t)$ 
transported along fluid streamlines in $M$, 
the entropy circulation integral \eqref{entropycircconslawSdimodd} is given by
\begin{equation}
\fr{d}{dt}\int_{\S(t)} f(\circ/\rho) g_{ij} \vortvec_{\S}^i \nabla_{\S}^j S \;dV_{\S}
= 0 , 
\end{equation}
where $\circ$ is the entropy circulation scalar \eqref{Mentrcirc} 
of the manifold $M$ when the dimension $n$ is odd,  
while for any orientable even-dimensional surface $\S(t)$, 
the entropy circulation integral \eqref{entropycircconslawSdimeven} is given by
\begin{equation}
\fr{d}{dt}\int_{\S(t)} f(S) \vortscal_{\S} \;dV_{\S}
= q\int_{\p\S(t)} e_S f(S) g_{ij} \vortvec_{\p\S}^i \nabla_{\p\S}^j S \;dA_{\S} 
\end{equation}
in terms of the projections of the gradient $\nabla^i$ 
into the tangent spaces of $\S$ and $\p\S$. 

In all of these integrals, 
\begin{equation}
dV_{\S}=\epsilon_{j_{1}\cdots j_{s}}(\S)  dx^{j_{1}}\wedge\cdots\wedge dx^{j_{s}}
\label{dV}
\end{equation}
is the volume element of the surface $\S$,
and 
\begin{equation}
dA_{\S}=\hat n^i\epsilon_{ij_{1}\cdots j_{s-1}}(\S) dx^{j_{1}}\wedge\cdots\wedge dx^{j_{s-1}}
\label{dA}
\end{equation}
is the volume element of the boundary $\p\S$,
with $s=\dim\S$, 
where $\epsilon_{j_{1}\cdots j_{s}}(\S)$ 
denotes the metric-normalized volume form of $\S$
and $\hat n^i$ denotes the unit normal of $\p\S$ in $\S$.

\end{document}